\documentclass[prl,aps,floats,amssymb,twocolumn,preprintnumbers,footnoteinbib]{revtex4}

\usepackage{epsfig}

\newcommand{\be}{\begin{equation}}
\newcommand{\ee}{\end{equation}}
\newcommand{\ba}{\begin{array}{c}}
\newcommand{\ea}{\end{array}}
\newcommand{\bqa}{\begin{eqnarray}}
\newcommand{\eqa}{\end{eqnarray}}

\begin{document}

\title{ Study on $X(3872)$ from effective field theory with pion exchange interaction}

\author{P.~Wang$^{a,b}$}
\author{X.~G.~Wang$^{a,b}$}

\affiliation{$^a$Institute of High Energy Physics, CAS, P. O. Box
918(4), Beijing 100049, China}

\affiliation{$^b$Theoretical Physics Center for Science Facilities,
CAS, Beijing 100049, China}

\begin{abstract}
We study $D\bar{D}^*$ ($D^*\bar{D}$) scattering in the framework of unitarized heavy meson chiral perturbation theory with pion exchange and a contact interaction.
$^3S_1-$$^3D_1$ mixing effects are taken into account. A loosely bound state $X(3872)$, with the pole position being $M_{pole}=(3871.70-i0.39)\mathrm{MeV}$, is found. The result is not sensitive to the strength of the contact interaction.
Our calculation provides a theoretical confirmation of the existence of the $1^{++}$ state $X(3872)$.
The light quark mass dependence of the pole position indicates it has a predominately $D\bar{D}^*$ ($D^*\bar{D}$) molecular nature. When the $\pi$ mass is larger than
142 MeV, the pole disappears which makes impossible the lattice simulation of this state at large quark mass.
\end{abstract}

\maketitle

The narrow resonance structure named $X(3872)$, discovered by the Belle Collaboration in the $B^+\rightarrow K^+J/\Psi\pi^+\pi^-$
process~\cite{Belle:03} and then confirmed by CDF and D0 Collaborations through its inclusive production in proton-antiproton
collisions~\cite{CDF:04,D0:04}, has inspired heated discussions both experimentally and theoretically.
In 2006, the Belle collaboration studied the $B^+\rightarrow D^0\bar{D}^0\pi^0K^+$ decay process and found an enhancement of
the $D^0\bar{D}^0\pi^0$ signal just above the $D^0\bar{D}^{*0}$ threshold~\cite{Belle:2006}, with the resonance is peaked at
\begin{equation}
M_X=3875.2\pm0.7^{+0.3}_{-1.6}\pm0.8\mathrm{MeV}\ .
\end{equation}
A later analysis of Belle data including new data on $D^*\rightarrow D\gamma$~\cite{Belle:2010} gave
\begin{equation}
M_X=3872.6^{+0.5}_{-0.4}\pm0.4\mathrm{MeV}\ .
\end{equation}
The latest PDG value for the $X(3872)$ mass from the $J/\Psi X$ decay mode is~\cite{PDG:2012}
\begin{equation}
M_X=3871.68\pm0.17\mathrm{MeV}\ .
\end{equation}
The angular distributions and correlations of the $\pi^+\pi^-J/\psi$ final state studied by the CDF collaboration indicate
two possible quantum numbers of this state, $J^{PC}=1^{++}$ or $2^{-+}$. The radiative
decay reported by Belle and BaBar bolsters the $1^{++}$ assignment, while the 3$\pi$ invariant mass
distribution in $J/\psi\omega$ decays slightly favors $2^{-+}$.

The line shapes of $B^+\rightarrow XK^+$ in the $J/\Psi\pi^+\pi^-$ and $D^0\bar{D}^0\pi^0/D^0\bar{D}^{*0}$ modes and
the corresponding pole structures have been studied independently by two groups~\cite{Hanhart:07,Braaten}. The effect of energy resolution effect was taken into account in~\cite{Zhang:09} and a twin-pole structure was found, suggesting that the $X(3872)$ can be identified as a $2^3P_1$ $c\bar{c}$ state strongly distorted by coupled channel effects. The effects on the lineshape of $X(3872)$ from nonzero decay width of $D^{*0}$ and inelastic channels was taken into account by Braaten~\cite{Braaten:09}.

The proximity of the $X(3872)$ to the threshold of $D^0\bar{D}^{*0}$ strongly suggests that the $X(3872)$ is probably a loosely bound $D^0\bar{D}^{*0}$ molecular state~\cite{Tornqvist:1994,Swanson:04,Thomas:2008,Liu:2009}. Other interpretations include normal charmonium~\cite{Barnes:04,Suzuki:2005,Meng:2007}, a tetraquark state~\cite{Maiani:05} and a $c\bar{c}g$ hybrid~\cite{Li:05}. In the above investigations, many kinds of potential models were used, leading to different conclusions. Recently, one-pion exchange as a possible binding mechanism in the $X(3872)$ was revisited \cite{Kalashnikova:12}.
The authors argued that it is not sufficiently binding for this purpose, suggesting other short-range interactions should be included and may be responsible for the $X(3872)$ formation.

Most of the theoretical studies can only provide a model and parameter dependent prediction for the $X(3872)$. Until now the theoretical situation concerning the existence of $X(3872)$ and its $J^{PC}$ number was unclear. Even for
lattice simulations, there are also many lattice studies predicting $1^{++}$ $X(3872)$ mass ranging from 3850 to 4060
MeV \cite{Okamoto,Chen1,Chen2,Liu}, but with various uncertainties of their own, where the key difficulty is the challenging task of extracting the excited states.
The lattice simulation indicates a lower mass with $J^{PC}$ number $2^{-+}$ \cite{Liu,Yang}.

Effective field theory is a very powerful tool to study hadron properties at low energy.
A pionless effective field theory describing $D\bar{D}^*$ scattering was proposed in~\cite{Alfiky:06},
where the $s$-channel bubble diagrams are summed by solving Lippmann-Schwinger equations to produce a bound state.
However, to obtain a pole near threshold, the low energy constants must be huge. In other words, for a small
``natural" interaction, there is no $1^{++}$ $X(3872)$ appears. The pion exchange interaction is well known from
the $D^*$ to $D\pi$ decay. We will see that with pion interaction alone, there is a bound state pole near threshold which is not sensitive
to the strength of the contact interaction.

In this Letter, we provide a reliable, parameter free calculation using unitarized chiral perturbation theory.
This approach has been successfully applied in the light and strange quark sector \cite{Dai}. It has also been applied to the heavy meson
case with open charm \cite{Guo:09,Wang:12}. The key advantage of this method is that the Lagrangian is well defined. The mass and width of the resonance or
bound state are obtained from the pole analysis of the scattering amplitude. One need not assume the constituent components
of the resonance as in potential models.

Here, we calculate $D\bar{D}^*$ ($D^*\bar{D}$) scattering amplitudes up to one loop in the framework of heavy meson chiral perturbation theory. We will focus on the $J^{PC}=1^{++}$ channel. The effects of $D^{*}$ finite width and $\ ^3S_1-$$^3D_1$ mixing can be naturally taken into account. Pad\'e approximation is used to construct the physically unitarized $T$ matrix.

We first construct $C$-parity even initial and final states, which are a superposition of $D^* \bar{D}$ and $D\bar{D}^*$,
\begin{equation}
|X_{+}>=\frac{1}{\sqrt{2}}(|D^* \bar{D}> + |D\bar{D}^*>)\ .
\end{equation}
We shall look for possible bound state, virtual state or resonant poles of the following transition amplitude,
\begin{equation}\label{T++}
T_{++}=\langle X_+|\hat{T}|X_+\rangle=\frac{1}{2}(T_{11}+T_{12}+T_{21}+T_{22})\ ,
\end{equation}
where
$T_{ij}$ are the four relevant transition amplitudes.
The general Lagrangian describing 4-boson contact interactions can be written as~\cite{Alfiky:06}
\begin{eqnarray}
&&\mathcal{L}^{(0)}\nonumber\\
&=&C_2 \left[ P^{(Q)\dag} P^{(\bar{Q})} V_{\mu}^{(\bar{Q})\dag} V^{(Q)\mu}+ P^{(\bar{Q})\dag} P^{(Q)} V_{\mu}^{(Q)\dag} V^{(\bar{Q})\mu}\right]\nonumber\\
&-&C_1 \left[ P^{(Q)\dag} P^{(Q)} V_{\mu}^{(\bar{Q})\dag} V^{(\bar{Q})\mu} + P^{(\bar{Q})\dag} P^{(\bar{Q})} V_{\mu}^{(Q)\dag} V^{(Q)\mu} \right] ,\nonumber\\
\end{eqnarray}
where $P^{(Q)}=(D^0,D^+,D_s^+)$ and $V^{(Q)}=(D^{*0},D^{*+},D^{*+}_s)$ are the heavy meson fields, while $P^{(\bar{Q})}=(\bar{D}^0,D^-,D_s^-)$ and $V^{(\bar{Q})}=(\bar{D}^{*0},D^{*-},D^{*-}_s)$ are the heavy antimeson fields.\\
\indent In order to include the effect from pion exchange on the properties of $X(3872)$, one needs to introduce the covariant chiral lagrangian describing interactions between heavy mesons and Goldstone bosons, which can be written as~\cite{Burdman:1992gh,Wise:1992hn,Yan:1992gz,Ding:09}
\begin{eqnarray}
\mathcal{L}^{(1)}&=&2g_{\pi} (V_{a\mu}^{(Q)\dag} P_b^{(Q)} + P_a^{(Q)\dag} V_{b\mu}^{(Q)}) u_{ba}^{\mu}\nonumber\\
&-& 2g_{\pi} (V_{a\mu}^{(\bar{Q})\dag} P_b^{(\bar{Q})} + P_a^{(\bar{Q})\dag} V_{b\mu}^{(\bar{Q})}) u_{ab}^{\mu} ,
\end{eqnarray}
where
\begin{equation}
u_{\mu}=i(u^{\dag}\partial_{\mu}u-u\partial_{\mu}u^{\dag})\ , \ \ \ \ u=\exp(\frac{i\phi}{\sqrt{2}F})\ ,
\end{equation}
with $\phi$ is the $3\times3$ matrix containing the Goldstone boson fields.
$F$ is the Goldstone boson decay constant in the chiral limit, which we identify with the pion decay constant, $F=92.4\mathrm{MeV}$. The coupling constant $g_{\pi}$ depends on heavy meson masses, $g_{\pi}=g\sqrt{M_D M_{D^*}}$ , where the dimensionless constant $g$ can be determined from the strong decay $D^{*+}\rightarrow D^+\pi^0$. Taking the rate from PDG~\cite{PDG:2012}, we find $g=0.30\pm0.03$.\\

The tree level scattering amplitudes from contact interactions have already been given in Ref.~\cite{Alfiky:06}.
\begin{figure}[h]%
\begin{center}%
 \mbox{\epsfxsize=60mm\epsffile{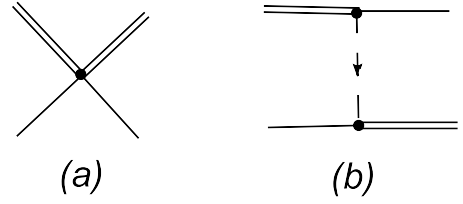}}%
\end{center}%
\caption{\label{TreeLevel} Tree level Feynman diagrams. Double lines indicate the vector $D^{*0}$ or $\bar{D}^{*0}$ mesons, solid lines pseudoscalar $D^0$ or $\bar{D}^0$ mesons. The dashed line represents the pion.}
\end{figure}
Tree level amplitudes from one-pion exchange should be added to the contact contributions, since both of them are of $\mathcal{O}(p^0)$,
\begin{eqnarray}
T_{11}&=&-C_1 \epsilon^{*}(p_3)\cdot\epsilon(p_1)\nonumber\\
T_{12}&=&C_2 \epsilon^{*}(p_3)\cdot\epsilon(p_1)+\frac{4g_{\pi}^2}{F^2}\frac{\epsilon^{*}(p_3)\cdot p_2 \epsilon(p_1)\cdot p_4}{u-M_{\pi}^2}\nonumber\\
T_{21}&=&C_2 \epsilon^{*}(p_3)\cdot\epsilon(p_1)+\frac{4g_{\pi}^2}{F^2}\frac{\epsilon^{*}(p_3)\cdot p_2 \epsilon(p_1)\cdot p_4}{u-M_{\pi}^2}\nonumber\\
T_{22}&=&-C_1 \epsilon^{*}(p_3)\cdot\epsilon(p_1)\ .
\end{eqnarray}
where
\begin{equation}
u(s,\cos\theta)=(p_1-p_4)^2=\Delta^2-p^2\frac{M_{D^*}^2+M_D^2}{M_{D^*}M_{D}}-2p^2\cos\theta\ ,
\end{equation}
with $\Delta=M_{D^*}-M_D$ and $p$ the external three momentum in the center-of-mass frame.\\
\indent In the following, we only focus on partial waves with $J=1$ and omit this index for simplicity. The projections of tree level amplitudes at $\mathcal{O}(p^0)$ with $l=l'=0$
and with positive $C$-parity are given by
\begin{equation}
T^{(0)}_{++,SS}=-(C_2-C_1)-\frac{2g_{\pi}^2}{3F^2}\left[2-\frac{\mu_{\pi}^2}{2p^2}\ln(1+\frac{4p^2}{\mu_{\pi}^2})\right]\ .
\end{equation}
where $\mu_{\pi}^2=m_{\pi}^2-\Delta^2$ is the effective pion mass. The superscript denotes the chiral order. In this work, we count $p, \mu_{\pi}\sim\mathcal{O}(p)$.\\

\begin{figure}[h]%
\begin{center}%
 \mbox{\epsfxsize=80mm\epsffile{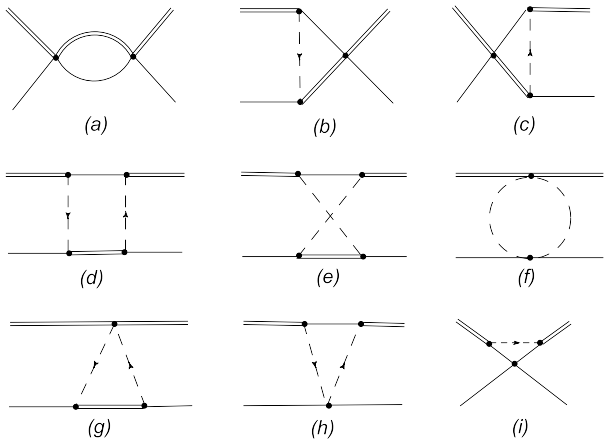}}%
\end{center}%
\caption{\label{one-loop} One loop Feynman diagrams. The notations are the same as Fig.~\ref{TreeLevel}. }
\end{figure}
One loop diagrams with contact and $D^*D\pi$ vertices are shown in Fig.~\ref{one-loop}, all of which are $\mathcal{O}(p^2)$ in a naive power counting scheme.
In the following, we use a similar procedure to that used in the analysis of $NN$ scattering~\cite{Kaiser:1997} to deal with the one loop diagrams. We will find that diagrams (a), (b), (c) and (d) start to contribute already at $\mathcal{O}(p)$.
The partial wave amplitudes have the usual unitarity cut along the positive real axis in the complex $p^2$ plane, starting at $p^2=0$, which is related to the s-channel $D^*\bar{D}$ intermediate states. We finally get
\begin{equation}
T_{++,SS}^{(1),a}=i(C_1-C_2)^2\frac{p}{8\pi(M_D+M_{D^*})}
\end{equation}
The contribution to the elastic scattering amplitude from the triangle one loop diagram with three relativistic propagators is shown in Fig.~\ref{one-loop}(b).
Finally, this yields partial wave amplitudes with positive $C$-parity:
\begin{equation}
T_{++,SS}^{(1),b}=T_{++,SS}^{(1),c}=\frac{(C_2-C_1)g_{\pi}^2}{12\pi F^2(M_D+M_{D^*})}\left[2ip-\mu_{\pi}^2\Gamma_0(p)\right],
\end{equation}
The contribution to the elastic scattering amplitude from planar box diagram is shown as Fig.~\ref{one-loop}(d).
The other diagrams in Fig.~2, starting to contribute at $\mathcal{O}(p^2)$, are neglected in our numerical calculation.

The partial wave amplitudes satisfy the perturbative unitarity condition
\begin{equation}
\mathrm{Im}T^{(1)}_{++}=T_{++}^{(0)}\frac{p}{8\pi\sqrt{s}}T^{(0)*}_{++}
\end{equation}
where
$T_{++}^{(0)}$ and $T_{++}^{(1)}$ are the $^3S_1-$$^3D_1$ mixed amplitudes
from which we can construct the physical, unitary amplitude using Pad\'e approximation,
\begin{equation}
T_{++}^{phy}=T_{++}^{(0)}\cdot[T_{++}^{(0)}-T_{++}^{(1)}]^{-1}\cdot T_{++}^{(0)}
\end{equation}
With our conventions, the relationship between $S$ matrix and $T$ matrix is given by
\begin{equation}
S=1+i\frac{p}{4\pi\sqrt{s}}T^{phy}_{++}(p)\ .
\end{equation}

The physical masses of the scattering particles are taken from PDG~\cite{PDG:2012},
$M_D=1864.91,M_{D^*}=2006.98, m_{\pi^0}=134.98$(in $\mathrm{MeV}$).\\

We first set $\lambda=C_2-C_1=0$, finding that pion exchange alone is strong enough to form a bound state just below $D^*\bar{D}$ threshold. The pole position in the complex $p$-plane and corresponding pole mass are $p=(-15.46+i24.62)\mathrm{MeV}, M=(3871.70-i0.39)\mathrm{MeV}$,
respectively. Regardless of the small imaginary part, which is due to the finite width of its constituent, $\Gamma(D^*\rightarrow D\pi)$, our result is in good agreement with Ref.~\cite{Tornqvist:1994}. That the obtained mass is very close to the experimental mass of the $X(3872)$ is remarkable since we did not adjust any parameter. The only parameter $g$ is fixed
by the decay width of $D^*\rightarrow D\pi$. On comparison in Ref.~\cite{Alfiky:06}, in order to get the correct mass of the $X(3872)$, the particular value of $\lambda_R$ ($8.4\times 10^{-4}$ MeV$^{-2}$) is chosen, corresponding to our dimensionless parameter $\lambda=\lambda_RM_D M_{D^*}=3144$.
The obtained mass is very sensitive to $\lambda$ and the binding energy is proportional to $1/\lambda^2$. The binding energy decreases with the
increasing interaction strength $\lambda$. Especially at $\lambda=0$, the binding energy is infinite. This behavior is not physically reasonable.
Most important, due to the particular choice of $\lambda$, it is not convincing that the $X(3872)$ should exist.
In our work, we want to determine the existence of $X(3872)$ theoretically without using any information concerning the $X(3872)$.
With the pion exchange interaction included, the pole is automatically generated and the mass is close to the experimental data without adjusting any parameter.

\begin{figure}[h]%
\begin{center}%
 \mbox{\epsfxsize=80mm\epsffile{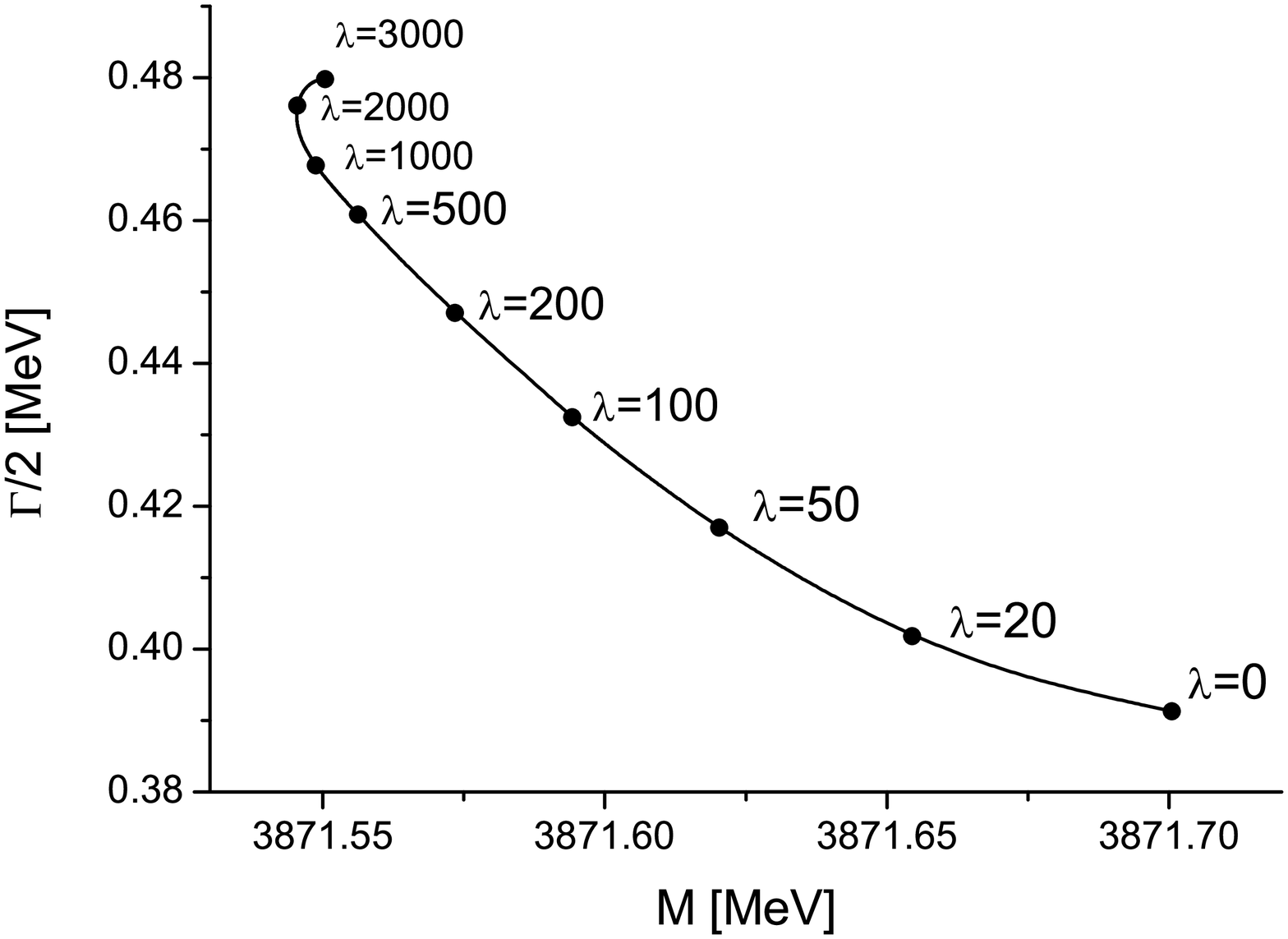}}%
\end{center}%
\caption{\label{pole-trajectory} Pole trajectory as increasing the strength of contact interaction.}
\end{figure}

We now check the effect of including additional attraction of shorter range by increasing the low energy constant $\lambda$.
The pole trajectory is shown in Fig.~\ref{pole-trajectory}, from which we can see that the binding energy changes slightly as we increase
the strength of the attractive contact interaction. For example, the mass changes from 3871.70MeV to 3871.55 MeV when $\lambda$ increases from $0$ to 3000.
It is clear that the $\lambda$ dependence of the pole is highly suppressed once pion exchange is included. It also shows that
the pion exchange interaction is the main reason for the system to be bound. Without the pion exchange interaction,
there is no pole around the $DD^*$ threshold with only a contact interaction except for the particular value of $\lambda$.
Our calculation theoretically confirms the weakly bound state $X(3872)$ with $J^{PC}$ number $1^{++}$ exists.
This is a prediction directly from the unitarized heavy meson chiral perturbation theory rather than fitting the experimental data.

Compared with potential models, the advantage of this pole analysis approach is that it does not specify
the components of the state. The obtained state is a physical state which could be a mixing of many pure states with the same $J^{PC}$ numbers.
In other words, we have no information of the structure of the state. For example, for $X(3872)$, we do not know whether the obtained state is
a traditional meson or tetraquark. Therefore, we need other methods to obtain the information concerning its structure.
As emphasized in a series of papers~\cite{Hanhart:08,Cleven:11}, the quark mass dependence can provide important
information on the structure of a state.
If the $X(3872)$ is a pure $c\bar{c}$ state with no constituent light quarks. Its light quark mass dependence only comes from sea quark
contributions, which should be very weak, as for the case of $D_s(1968)$ shown in lattice simulations \cite{Follana}.

\begin{figure}[h]%
\begin{center}%
 \mbox{\epsfxsize=80mm\epsffile{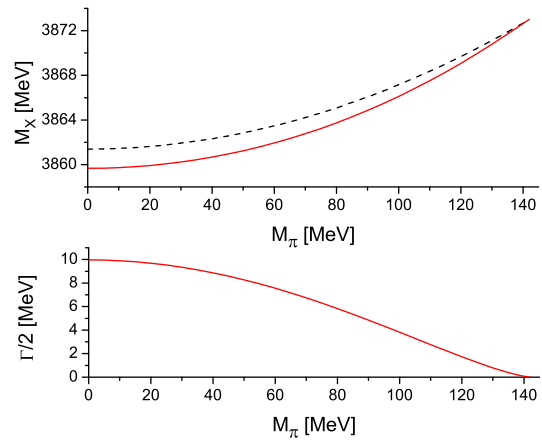}}%
\end{center}%
\caption{\label{Mx-mpi} Upper plot: Pion mass dependence of the mass of $X(3872)$ (solid line) and $D^*\bar{D}$ threshold (dashed line);
Lower plot: Pion mass dependence of $\Gamma_{X(3872)}/2$. }
\end{figure}

We fix the $c$ quark mass at its physical value and vary the light quark masses (pion mass). At tree level,
the pion mass dependence of  $D^{(*)}$ mesons can be expressed as \cite{Wang:12}
\begin{equation}
M_{D^{(*)}}(M_{\pi})=M_{D^{(*)}}|_{phy}+\frac{2h_0+h_1}{M_{D^{(*)}}|_{phy}}(M_{\pi}^2-M^2_{\pi}|_{phy})\ ,
\end{equation}
where the values of $h_0$ and $h_1$ are taken from~\cite{Wang:12}.
The pion mass dependence of the mass of $X(3872)$ as well as the threshold is shown in Fig.~\ref{Mx-mpi}.
From the figure, one can see that for a small change of quark mass (from zero to several MeV),
there is an obvious change of the $X(3872)$ mass which shows the light quark component in the state.

In general, the mass of a loosely bound $A-B$ molecular state is given by
\begin{equation}
M_b=M_A+M_B-E_b\ ,
\end{equation}
where $E_b$ is the binding energy. The pion mass dependence of $E_b$ is expected to be much weaker than that of $M_A$ and $M_B$.
Thus, the mass of an $A-B$ molecular state is almost the same as that of the $A-B$ threshold.
From Fig.~\ref{Mx-mpi}, one can see that the pion mass dependence of the $X(3872)$ is in good agreement with the expectation for a $D^*-\bar{D}$ molecular state.
We can also see that the tiny width decreases with increasing pion mass, since the phase space for $D^*$ decay to $D\pi$ will be suppressed.
Eventually, the bound state pole will be absorbed by the $D^*\bar{D}$ threshold as $\mu_{\pi}\rightarrow 0$.

When $m_\pi$ is large than $\Delta$ (142 MeV), there is no bound state or resonance pole.
This is an important result related to the lattice simulation. Recently, a lattice simulation
result showed that there is no $1^{++}$ state near the experimental mass of $X(3872)$ \cite{Liu}.
The authors explained that if $X(3872)$ was a molecule state, it was not expected to be found in their simulation.
Our result shows if the simulation is at large quark mass,
it certainly can not be found for any $1^{++}$ operator, even though the operator has a
large coupling with the molecule state.

In summary, we studied the $D\bar{D}^*$ ($D^*\bar{D}$) scattering using unitarized heavy meson chiral perturbation theory.
The standard one pion exchange is included in the chiral Lagrangian. The parameter $g$
is determined from the PDG value for the decay $D^*\rightarrow D\pi$. The obtained pole mass is
very close to the experimental data which is not sensitive to the low energy constant $\lambda$.
This confirmation of the existence of the $1^{++}$ $X(3872)$ is a parameter-free prediction rather than
a fitting of the experimental data. The light quark (pion) mass dependent behavior shows it is probably
a $D^*-\bar{D}$ molecule. The disappearance of the pole at large pion mass makes it impossible
for current lattice simulations to find the $1^{++}$ $X(3872)$ . Future lattice simulations must be carried out
at the physical pion mass in order to find it.

\section*{Acknowledgments}

P.W. is grateful to A. W. Thomas for helpful discussions.
This work is supported in part by DFG and NSFC (CRC 110) and by the
National Natural Science Foundation of China (Grant No. 11035006).

\end{document}